\AtBeginDvi{}

\documentclass[reprint,groupedaddress]{revtex4-1}
\usepackage[dvips]{graphicx}
\usepackage{amsthm}
\usepackage{amsmath}
\usepackage{physics}
\usepackage{here}
\usepackage{amssymb}
\usepackage{color}
\usepackage{bm}
\usepackage{bbold}
\usepackage{letltxmacro}
\usepackage{algorithm}
\usepackage{algorithmic}
\usepackage{threeparttable}
\usepackage{caption}
\usepackage[usenames,dvipsnames]{xcolor}
\usepackage[breaklinks,colorlinks=true,citecolor=blue,linkcolor=RubineRed,urlcolor=RubineRed]{hyperref}

\begin{document}
\title{Latency-aware adaptive shot allocation for run-time efficient variational quantum algorithms}

\author{Kosuke Ito}
\affiliation{Center for Quantum Information and Quantum Biology, International Advanced Research Institute, Osaka University, Osaka 560-8531, Japan}

\begin{abstract}
 Efficient classical optimizers are crucial in practical implementations of Variational Quantum Algorithms (VQAs). In particular, to make Stochastic Gradient Descent (SGD) resource efficient, adaptive strategies have been proposed to determine the number of measurement shots used to estimate the gradient. However, existing strategies overlook the overhead that occurs in each iteration. In terms of wall-clock runtime, significant circuit-switching and communication latency can slow the optimization process when using a large number of iterations. In terms of economic cost when using cloud services, per-task prices can become significant. To address these issues, we present an adaptive strategy that balances the number of shots in each iteration to maximize expected gain per unit time or cost by explicitly taking into account the overhead. Our approach can be applied to not only to the simple SGD but also its variants, including Adam. Numerical simulations show that our adaptive shot strategy is actually efficient for Adam, outperforming many existing state-of-the-art adaptive shot optimizers. However, this is not the case for the simple SGD. When focusing on the number of shots as the resource, our adaptive-shots Adam with zero-overhead also outperforms existing optimizers. 
\end{abstract}


\maketitle

\section{Introduction}
Variational quantum algorithms (VQAs) have attracted much attention as a promising candidate for the first practical use of noisy intermediate-scale quantum (NISQ) devices \cite{Preskill2018quantumcomputingin,Cerezo:2021wl}.
In VQAs, the cost function of a variational problem is computed by utilizing somehow shallow quantum circuits, and the optimization of the variational parameters is done on a classical computer.
Several paradigms of VQAs have been proposed, such as
the variational quantum eigensolver (VQE) \cite{Peruzzo:2014uz,PhysRevX.6.031007,PhysRevX.6.031045,Kandala:2017wb} to obtain an approximation of the ground-state energy of a Hamiltonian and beyond \cite{McClean2017,Santagatieaap9646,1810.12745,PhysRevX.8.011021,McArdle:2019we,PhysRevA.99.062304,Parrish2019,Higgott2019,PhysRevResearch.1.033062,tilly2020computation,Ollitrault2019},
the quantum approximate optimization algorithm (QAOA) \cite{1411.4028,1602.07674,1712.05771} for combinatorial optimization problems, quantum machine learning \cite{PhysRevA.98.032309,Benedetti_2019,2004.11280}, and many others \cite{Cerezo:2021wl}.

One of the key to successful VQAs is an efficient optimization strategy of the classical parameters.
To go beyond simple applications of existing non-quantum specific classical optimizers such as Adam, Nelder-Mead, Powell, etc.~\cite{9259985,Kubler2020adaptiveoptimizer,LaRose:2019tb}, several quantum-aware optimization algorithms have been proposed \cite{Kubler2020adaptiveoptimizer,2004.06252,Stokes2020quantumnatural,PhysRevA.106.062416,nakanishi_sequential_2020,1904.03206,Sweke2020stochasticgradient,PhysRevResearch.4.023017,PRXQuantum.2.030324,2108.10434,2201.13438,Tamiya:2022um,2210.06484,2211.04965,PhysRevX.12.041022}.
Especially, it have been observed that optimizers using gradient information offer improved convergence over those without it \cite{PhysRevLett.126.140502}.
A problem of gradient based optimizers is that the calculation of the gradient is expensive in VQAs since each component of the gradient must be individually evaluated.
In fact, the choice of the number of measurement shots to estimate the gradient severely affects the performance of gradient based optimizers \cite{Sweke2020stochasticgradient}.
Then, adaptive strategies to assign appropriate numbers of shots for efficient optimization have been proposed \cite{Kubler2020adaptiveoptimizer,2004.06252,2108.10434}.
Those strategies actually achieve fast convergence with respect to the total shot number.

However, in practice, the total time consumption is more important rather than the total number of shots.
The latency occurs with circuit-switching and the communication is usually much larger than the single-shot acquisition time \cite{Karalekas_2020,2201.13438}.
This situation is similar for economic cost of using a cloud quantum computing service, where per-task prices are usually much more expensive than single-shot prices \cite{aws:pricing}.
To our best knowledge, no optimizer has been proposed in such a way that the value of the latency time is explicitly reflected to the optimization strategy.
In this paper, we propose an approach for improving the efficiency of shot-adaptive gradient-based optimizers in terms of total time consumption or cost for using a cloud service. The approach, referred to as the Waiting-time Evaluating Coupled Adaptive Number of Shot (weCANS), determines the number of shots by evaluating the waiting time or per-task cost to optimize overall efficiency.
Our proposal includes a latency-aware adaptive-shots Adam, which we call we-AdamCANS, as well as simple generalizations of iCANS and gCANS. Our research aims to address the potential improvement in the performance of gradient-based optimizers through explicit consideration of latency values in their design.

This paper is organized as follows.
Sec.~\ref{juxulzhd} provides an overview of the background related to the topic.
Sec.~\ref{main_alg} introduces the weCANS algorithm. The section begins by providing a simple generalization of the iCANS and gCANS algorithms (in Sec.~\ref{weCANS1}), followed by a generalization of weCANS to a wider range of stochastic gradient descent methods, including the popular Adam algorithm (in Sec.~\ref{sec_AdamCANS}).
In Sec.~\ref{sec_weCANS-WRS}, we describe how to incorporate random operator sampling techniques for measuring Hamiltonians that consist of multiple noncommuting terms into the weCANS algorithm.
Sec.~\ref{sec_numerics} presents numerical simulations of the proposed strategies on various tasks including Compiling task, VQE tasks of quantum chemistry, and VQE of 1-dimensional Transverse field Ising model.
Finally, the conclusion of the paper is presented in Section~\ref{jzbmlzhd}.

\section{Backgrounds}
\label{juxulzhd}
\subsection{Variational quantum eigensolver}
\label{jvbilzhd}
As a typical VQA,
we focus on problems formulated as the minimization of the cost function
\begin{align}
 f(\bm{\theta}^{(k)}) := \bra{\psi(\bm{\theta})} H \ket*{\psi(\bm{\theta})}.\label{VQE-cost}
\end{align}
given by the expectation value of a Hermitian operator $H$.
In VQE \cite{Peruzzo:2014uz,PhysRevX.6.031045,Kandala:2017wb}, we attempt to find a good approximation of the ground state energy of a quantum mechanical Hamiltonian $H$ by minimizing the cost function \eqref{VQE-cost}.
It is expected that a quantum computer outperforms in computing the expectation value $\bra{\psi(\bm{\theta})} H \ket*{\psi(\bm{\theta})}$ of a quantum mechanical Hamiltonian of some complex systems.
Hence, a quantum advantage might be achieved by only using the quantum device to compute the expectation value, while the parameter vector $\bm{\theta}^{(k)}$ is classically optimized.
That is why the VQE is a candidate for a practical application of NISQ devices.
For a successful VQE, it is important to find an efficient classical optimizer as well as
an expressive and trainable ansatz $\ket{\psi(\bm{\theta})}$ which is realizable in a quantum computer.
Our focus in this paper is to find an efficient optimizer with respect to the total wall-clock running time and the economic cost as practical figures of merit.
Especially, we focus on the gradient descent optimizers.

One possible way to approximately estimate the partial derivative is the numerical differentiation:
\begin{align}
 \frac{\partial f}{\partial \theta_i} (\bm{\theta}) \approx \frac{f(\bm{\theta} + \delta \bm{e}_i) - f(\bm{\theta})}{\delta},
\end{align}
where $\bm{e}_i$ is the unit vector in the $i$-th direction and $\delta$ is a small number.
However, as the finite numbers of measurements for the estimations of the cost function values introduce statistical error, this estimation is not good \cite{PhysRevLett.126.140502}.
Instead, we can use an analytic expression of the gradient of the cost function in VQAs.
If the ansatz is given by parametric gates in the form of $U_i(\theta_i) = \exp [-i \theta_i A_i / 2]$ with $A_i^2 = I$ and fixed gates,
we can use the following so-called parameter shift rule \cite{PhysRevA.98.032309,PhysRevA.99.032331}:
\begin{align}
 \frac{\partial f}{\partial \theta_i} (\bm{\theta}) = \frac{1}{2}\left(f(\bm{\theta} + \frac{\pi}{2} \bm{e}_i) - f(\bm{\theta} - \frac{\pi}{2} \bm{e}_i)\right).\label{PSR_2}
\end{align}
For example, hardware efficient ansatz actually fits in this form.
More generally, the partial derivatives may be calculated by general parameter shift rules with some $R_i$ number of parameter shifts \cite{Wierichs2022generalparameter}:
\begin{align}
 \frac{\partial f}{\partial \theta_i} (\bm{\theta}) = \sum_{j=1}^{R_i} a_{j} f(\bm{\theta} + x_{i,j}\bm{e}_i).\label{GPSR}
\end{align}
Hence, the gradient can be computed by evaluating the cost function at shifted parameters and taking their linear combination.
We remark that the statistical error in the estimation of the cost function still remains due to the finiteness of the number of quantum measurements even if we use the analytic form.

\subsection{Operator sampling}\label{sec_op_sampling}
The target Hamiltonian $H$ often cannot be measured directly in a quantum computer (e.~g.~Hamiltonians in quantum chemistry).
In such a case, we decompose $H$ into directly measurable operators such as Pauli operators $P_i$ as
\begin{align}
 H = \sum_{i=1}^{N} c_i P_i.\label{pauli_decomp}
\end{align}
In general, we cannot measure all the terms simultaneously because this decomposition includes noncommuting Pauli operators.
To efficiently measure the observables $P_i$, we need to group simultaneously measurable observables.
Finding an efficient strategy to choose appropriate measurements is an important problem which have been intensively studied \cite{PhysRevX.6.031007,Kandala:2017wb,Hamamura:2020tc,Crawford2021efficientquantum,PhysRevX.8.031022,Izmaylov:2020wc,Vallury2020quantumcomputed,doi:10.1063/1.5141458,PhysRevA.101.062322,Wu2023overlappedgrouping}.
We do not go into details regarding the grouping problem in this paper.
Once the groups $G_j$ of the commuting operators are fixed, another issue is how to allocate the number of shots to each group under a given shot budget $s_{\mathrm{tot}}$.
Although this is also a nontrivial problem, we take the weighted random sampling (WRS) strategy \cite{2004.06252} in this paper.
In WRS, each $j$-th group is randomly selected with probability $\sum_{i\in G_j}|c_i| / \sum_{k=1}^N |c_k|$ in each measurement.
We can implement this procedure by sampling the number $s^{(j)}$ of shots for each $j$-th group from the multinomial distribution for $s_{\mathrm{tot}}$ independent trials with probability distribution $p_j = \sum_{i\in G_j}|c_i| / \sum_{k=1}^N |c_k|$.
If we allocate deterministic number of shots, some groups may never be measured for small $s_{\mathrm{tot}}$, which results in the bias in the estimation of the expectation value of $H$.
In WRS, we obtain an unbiased estimator of the expectation value of $H$ for any small number of shots $s_{\mathrm{tot}}$.

\subsection{Stochastic gradient descent}\label{BG_SGD}
Gradient descent is a very standard approach to optimization problems.
In the gradient descent, we minimize the cost function $f(\bm{\theta})$ by iteratively updating the parameter vector $\bm{\theta} \in \mathbb{R}^d$ as
\begin{align}
 \bm{\theta}^{(k+1)} = \bm{\theta}^{(k)} - \alpha \bm{\nabla} f(\bm{\theta}^{(k)}),\label{GD-eq1}
\end{align}
where $\alpha$ is the learning rate which is a hyperparameter.
In some problems such as machine learning tasks, $\bm{\nabla} f(\bm{\theta}^{(k)})$ in \eqref{GD-eq1} can be replaced with an estimator $\bm{g}(\bm{\theta}^{(k)})$ of it, which is a random variable.
Gradient descent using a random estimator of the gradient is called stochastic gradient descent (SGD).
When the estimator is unbiased, rigorous convergence guarantees are available under appropriate conditions \cite{shalev2014understanding,10.1007/978-3-319-46128-1_50,Sweke2020stochasticgradient}.
Moreover, the stochasticity can rather help to avoid local minima and saddle points \cite{Bottou1991StochasticGL}.
Thus, SGD is often efficient not only because of saving the resource to compute the gradient.

In VQAs, we have no access to the full gradient $\bm{\nabla} f$ due to the essentially random nature of the quantum measurement.
Hence, the gradient descent in VQAs is inevitably stochastic in some extent depending on the number of measurements used to estimate the gradient.
Although using large numbers of measurements results in accurate estimation of the full gradient, the accurate estimation of the gradient is not necessarily beneficial, as the stochasticity can be rather advantageous in SGD.
In fact, it is reported that using not large numbers of measurements for gradient evaluations effectively results in the implementation of SGD \cite{PhysRevLett.126.140502}, and is actually beneficial in the optimization \cite{Sweke2020stochasticgradient,2210.06723}.
However, the accuracy of the estimation of the gradients also restricts the final precision of the optimization.
Therefore, the numbers of measurements for evaluation of the gradients during the optimization is an important hyperparameter which determines the effectiveness of the optimizer.
While the numbers of measurements as well as the learning rate are usually tuned heuristically, some papers propose algorithms to adaptively set the numbers of measurements in SGD as we review later in more detail (Sec.~\ref{sec_adaptive}) \cite{Kubler2020adaptiveoptimizer,2004.06252,2108.10434}.
The adaptive allocation of appropriate numbers of measurement shots for practically efficient SGD is the main focus of our study.

For the selection of the learning rate $\alpha$, we can make use of the Lipschitz continuity of the gradient of the cost function of the VQE:
\begin{align}
 \|\bm{\nabla}f(\bm{\theta}^{(k+1)}) - f(\bm{\theta}^{(k)})\|\leq L \|\bm{\theta}^{(k+1)} - \bm{\theta}^{(k)}\|,\label{Lips_conti}
\end{align}
where $L$ is called a Lipschitz constant.
According to Eq.~(\ref{Lips_conti}),
we can make the change in the gradient per single descent step small by taking small enough $\alpha$ in accordance with the size of $L$.
In this way, it is expected that the course of the optimization well follows the gradient.
In fact, for the exact gradient descent, the convergence is guaranteed if $\alpha < 2/L$ \cite{Balles:2017ta}.
As seen later, the Lipschitz constant is also used to set the number of shots in our algorithm in a similar manner to \cite{Kubler2020adaptiveoptimizer}. 
For the cost function of the VQE, we have access to a Lipschitz constant.
Especially, for the case with the parameter shift rule Eq.~(\ref{PSR_2}), we see that any order of the partial derivatives of $f$ is upper bounded by the matrix norm $\|H\|$ of $H$ by recursively applying Eq.~(\ref{PSR_2}).
Then, if the parameter vector is $d$-dimensional, we have the following upper bound of the Lipschitz constant \cite{Sweke2020stochasticgradient,2108.10434} 
\begin{align}
 L \leq d \|H\|.\label{Lips_const}
\end{align}
When $H$ is decomposed as (\ref{pauli_decomp}), the norm is bounded as $\|H\|\leq \sum_{i=1}^{N} |c_i|$.
We remark that smaller estimation of $L$ results in good performance of optimizers using $L$ in general.
If we need tighter estimation of $L$ than \eqref{Lips_const}, we can search for smaller $L$ in a similar manner to hyperparameter tuning.

\subsection{Adam}\label{sec_Adam}
Adaptive moment estimation (Adam) \cite{Kingma:2014aa} is a variation of a modified SGD which adaptively modifies the descent direction and the step size using the estimations of first and the second moments of the gradient.
In Adam, we use the exponential moving averages
\begin{align}
 \bm{m}_{k} = \beta_1 \bm{m}_{k-1} + (1 - \beta_1) \bm{g}(\bm{\theta}^{(k)})\label{adam_m}
\end{align}
\begin{align}
 \bm{v}_{k} = \beta_2 \bm{v}_{k-1} + (1 - \beta_2) \bm{g}(\bm{\theta}^{(k)})^2\label{adam_v}
\end{align}
as the estimations of the moments, where the multiplication of the vector is taken component-wise, $\beta_1, \beta_2$ are the hyperparameters which determines the speed of the reflection of the change in the gradient.
Then, we update the parameter according to
\begin{align}
 \bm{\theta}^{(k+1)} = \bm{\theta}^{(k)} - \alpha_k \frac{\bm{m}_k}{\sqrt{\bm{v}_k} + \epsilon},\label{adam_update}
\end{align}
where $\alpha_k$ is the step size which depends on $k$ in general, and $\epsilon$ is a small constant for the numerical stability.
The bias-corrected version of $\bm{m}_k$ and $\bm{v}_k$ and constant $\alpha_k$ are used in the original proposal of Adam, which is equivalent to using $\alpha_k = \alpha \sqrt{1 - \beta_2^k} / (1 - \beta_1^k)$ with a constant $\alpha$ if we neglect $\epsilon$.
A possible advantage of Adam is that the step size for each component is adjusted component-wise, which gives an additional flexibility.
In fact, Adam is known to work well for many applications.
In this paper, we investigate how to exploit Adam's power in VQAs by using adaptively determined appropriate number of shots in Sec.~\ref{sec_AdamCANS}.

\subsection{Shot adaptive optimizers}\label{sec_adaptive}
As explained in Sec.~\ref{BG_SGD}, the number of shots for the estimation of the gradient is an important hyperparameter of SGD for VQE.
Using too large number of shots not only results in too expensive optimization, but also loses the benefit from the stochasticity.
On the other hand, too small number of shots results in the fail of the accurate optimization due to the statistical errors.
In general, the required number of shots tends to be small in the early steps in SGD, and the required number increases as the optimization progresses.
Thus, it is hard for SGD to be efficient if we give a fixed number of shots used throughout the optimization process.
One option to circumvent this problem is to dynamically increase the number $s$ of shots following a fixed schedule, e.~g.~$s = \lfloor s_0 r^k \rfloor$ with the initial number $s_0$ and a hyperparameter $r$.
However, very careful hyperparameter tuning is needed for such a strategy to be efficient \cite{2108.10434}.
Hence, an adaptive method to determine appropriate numbers of shots step by step in SGD optimizations is desired.

Balles et al.~\cite{Balles:2017ta} introduced Coupled Adaptive Batch Size (CABS) method for choosing the sample size in the context of optimizing neural networks by SGD.
Inspired by CABS, K{\"{u}}bler et al.~\cite{Kubler2020adaptiveoptimizer} proposed individual-Coupled Adaptive Number of Shots (iCANS) for choosing the number of shots for SGD optimizations in VQAs.
Both algorithms assign the sample size so that the expected gain per single sample is maximized.
To do so, the expectation value of a lower bound of the gain is considered based on the following inequality:
\begin{align}
 &f(\bm{\theta}^{(k)}) - f(\bm{\theta}^{(k+1)}) \nonumber\\
\geq& \alpha \bm{\nabla} f(\bm{\theta}^{(k)})^{\mathrm{T}} \bm{g}(\bm{\theta}^{(k)}) - \frac{L\alpha^2}{2}\|\bm{g}(\bm{\theta}^{(k)})\|^2 =: \mathcal{G},
\end{align}
where $\bm{g}(\bm{\theta}^{(k)})$ is an estimator of the gradient.
For VQAs, when each $i$-th component of the gradient is estimated by $s_i$ shots,
the expectation value of this lower bound $\mathcal{G}(\bm{s})$ reads
\begin{align}
 \mathbb{E}[\mathcal{G}(\bm{s})] = \left(\alpha - \frac{L\alpha^2}{2}\right)\|\bm{\nabla}f(\bm{\theta}^{(k)})\|^2 - \frac{L\alpha^2}{2}\sum_i \frac{\sigma_i^2}{s_i},
\end{align}
where $\bm{s}$ denotes $(s_1,s_2,\cdots,s_{d})$, and $\sigma_i$ is the standard deviation of $i$-th component $g_i(\bm{\theta}^{(k)})$ of the random estimator of the gradient.
We remark that to guarantee that $\mathbb{E}[\mathcal{G}]$ is positive, we need \cite{Balles:2017ta}
\begin{align}
 \alpha \leq \frac{2\|\bm{\nabla}f(\bm{\theta}^{(k)})\|^2}{L(\|\bm{\nabla}f(\bm{\theta}^{(k)})\|^2 + \sum_i \frac{\sigma_i^2}{s_i})}.
\end{align}
The gain $\mathcal{G}(\bm{s})$ can be divided into a sum of the individual contributions $\mathcal{G}_i(s_i)$ from each $i$-th component given as
\begin{align}
 \mathbb{E}[\mathcal{G}_i(s_i)] = \left(\alpha - \frac{L\alpha^2}{2}\right)\partial_i f(\bm{\theta}^{(k)})^2 - \frac{L\alpha^2}{2} \frac{\sigma_i^2}{s_i},
\end{align}
where $\partial_i f := \frac{\partial f}{\partial \theta_i}$.
In iCANS, we assign a different number $s_i$ of shots for estimating each component of the gradient by individually maximizing $\gamma_i := \mathbb{E}[\mathcal{G}_i(s_i)] / s_i$.
Then, the maximization of $\gamma_i$ yields the suggested number of shots \cite{Kubler2020adaptiveoptimizer}
\begin{align}
 s_i = \frac{2L\alpha}{2 - L\alpha}\frac{\sigma_i^2}{\partial_i f(\bm{\theta}^{(k)})^2}.\label{iCANS_si}
\end{align}
We have to note that this formalism is only valid when $\alpha < 2/L$.
The suggested number of shots implies that the larger number of shots should be assigned as the statistical noise $\sigma_i^2$ gets larger compared to the signal $\partial_i f(\bm{\theta}^{(k)})$.
Hence, $s_i$ tends to become large as the optimization progresses.
Moreover, a restriction on the number of shots not to exceed that of the highest expected gain is introduced in iCANS for further shot frugality.
That is, we impose $s_i \leq s_{\max}$ with the number $s_{\max}$ given by
\begin{align}
 i_{\max} &= \arg \max_i (\gamma_i)\\
 s_{\max} &= s_{i_{\max}}.
\end{align}
Fortunately, a Lipschitz constant $L$ to calculate $s_i$ is accessible in VQAs from Eq.~\eqref{Lips_const}.
However, the quantities $\sigma_i^2$ and $\partial_i f(\bm{\theta}^{(k)})$ are not accessible because $s_i$ must be calculated before we estimate the gradient.
Then, in order to implement iCANS, we replace these quantities with accessible quantities calculated from the previous estimations since it is expected that the change in these quantities along the course of SGD is not so large.
For increased stability, it is suggested to use bias-corrected exponential moving averages of the previous iterations in place of $\sigma_i^2$ and $\partial_i f(\bm{\theta}^{(k)})$.
We also employ this strategy in our algorithm.

Gu et al. \cite{2108.10434} proposed another algorithm global-CANS (gCANS) with a different rule to allocate the number of shots as a variant of CANS.
In gCANS, the rate of the expected gain is maximized globally, whereas each individual component is maximized in iCANS.
In other words, the number of shots $s_i$ is determined so that
\begin{align}
 \gamma = \frac{\mathbb{E}[\mathcal{G}(\bm{s})]}{\sum_{i=1}^{d}s_i}
\end{align}
is maximized.
Then, the suggested rule is
\begin{align}
 s_i = \frac{2L\alpha}{2 - L\alpha}\frac{\sigma_i\sum_{j=1}^{d}\sigma_j}{\|\bm{\nabla}f(\bm{\theta}^{(k)})\|^2}.
\end{align}
Inaccessible quantities are also replaced by the exponential moving averages in gCANS.
Both iCANS and gCANS have shown its efficiency in terms of the total number of shots used as a figure of merit \cite{Kubler2020adaptiveoptimizer,2108.10434}, and according to Ref.~\cite{2108.10434}, gCANS performs better than iCANS.
However, in practice, the total time consumption is more important rather than the total number of shots.
The latency occurs with circuit-switching and the communication is usually much larger than the single-shot acquisition time \cite{Karalekas_2020,2201.13438}.
As for the economic cost for using a cloud quantum computing service, the task cost is usually much more expensive than a single-shot cost \cite{aws:pricing}.
Although Ref.~\cite{2108.10434} also showed that gCANS has high frugality of task cost of Amazon braket pricing and the number of the iterations, they does not explicitly consider the performance with respect to the total time or the total economic cost taking into account the latency or task cost.

On the other hand, Menickelly et al.~\cite{2201.13438} investigated the performance of the optimizers of VQAs in terms of the total time including the latency.
They proposed a new optimizer, SHOt-Adaptive Line Search (SHOALS).
SHOALS makes use of stochastic line search based on \cite{Paquette:2020wb,Berahas:2021uq,NEURIPS2021_4cb81113} for the better iteration complexity.
In SHOALS, the number of shots used for estimating the gradient is adaptively determined so that the error in the estimations will be bounded in such a way that it does not interfere with the gradient descent dynamics.
The number of shots used for the cost function evaluations for the line search is also adaptively assigned in a similar manner.
The suggested number of shots is also basically proportional to the ratio of the noise to the signal, though the gain is not taken into account.
They have shown that SHOALS actually achieves better iteration complexity scaling $O(\epsilon^{-2})$ compared to that of the simple SGD $O(\epsilon^{-4})$, where $\epsilon$ is the precision to be achieved in expectation.
In that sense, they suggest SHOALS as a high-performing optimizer in terms of the total time in the presence of the large latency.
However, SHOALS does not take into account the explicit value of the latency.
To our best knowledge, no optimizer has been proposed in such a way that the value of the latency is explicitly reflected to the optimization strategy.
Therefore, a question arises here as to what extent the performance of an optimizer in terms of the total time consumption (or the economic cost) is improved by designing it with an explicit reflection of the latency value.

\section{Waiting-time evaluating coupled adaptive number of shots}\label{main_alg}
In this section, we propose adaptive shot allocation strategies taking into account the latency or per-task cost as the overhead.
We follow Ref.~\cite{2201.13438} for the treatment of the latency.
We denote the single-shot acquisition time by $c_1$, the circuit-switching latency by $c_2$, and the communication latency by $c_3$.
The single-shot acquisition time includes the time for the initialization, running a given circuit, and a single measurement.
The values of $c_1$ for example are around $100$ $\mu$s in superconducting qubits \cite{Karalekas_2020} and around $200$ ms in neutral atom systems \cite{doi:10.1080/09500340008244052}.
The circuit switching latency includes the time for compiling a circuit and load it on the controller of the quantum device.
For superconducting qubits, the compilation task takes $200$ ms, and interaction with the arbitrary waveform generator takes around $25$ ms \cite{Karalekas_2020,2201.13438}.
The communication latency occurs on the communication between the classical computer and the quantum device side.
Its value depends on the environment using the quantum computer, e.~g.~very small when we directly use a quantum device in the same laboratory, or many seconds when we communicate with a quantum device through the internet \cite{Sung_2020,2201.13438}.
The task cost overhead for the economic cost of using a cloud quantum computing service can also be treated in the same way, where $c_1$ is the per-shot price and $c_2$ is the per-task price and $c_3=0$.
For example, $c_1 = \text{\$} 3.5\times 10^{-4}$ and $c_2= \text{\$} 0.3$ for the pricing of the Rigetti's devices in Amazon Braket \cite{aws:pricing}.

\subsection{Simple latency-aware generalization of iCANS and gCANS}\label{weCANS1}
The simplest idea to take into account the latency is to modify the shot allocation rules of iCANS and gCANS in such a way that we maximize the gain per unit time (or economic cost) including the latency instead of the gain per single shot.
In other words, to modify iCANS to take into account the latency overhead, we assign the number of shots $s_i$ to estimate $i$-th component of the gradient such that
\begin{align}
 r_{\mathrm{i}}(s_i) = \frac{\mathbb{E}[\mathcal{G}_i(s_i)]}{c_1 s_i + c_2 m_i + c_3/d}\label{iCANS_max}
\end{align}
is maximized, where $m_i$ is the number of circuits used to estimate $i$-th component of the gradient.
For the case where $m_i$ is a random variable as is the case with WRS, we need to replace $m_i$ with some available estimated quantity.
We deal with this problem in detail in Sec.~\ref{sec_weCANS-WRS}.
Here, we assume that a single round of the communication is enough to estimate all the components of the gradient.
This maximization yields the optimal number of shots as
\begin{align}
 &s_i \nonumber\\
=& \frac{L\alpha}{2 - L\alpha}\left(1 + \sqrt{1 + R_i \frac{2 - L\alpha}{L\alpha} \frac{\partial_i f(\bm{\theta}^{(k)})^2}{\sigma_i^2}}\right)\frac{\sigma_i^2}{\partial_i f(\bm{\theta}^{(k)})^2},\label{weCANSi_shots}
\end{align}
where $R_i := (c_2 m_i + c_3 / d)/c_1$ is the ratio of the overhead per evaluation of each $i$-th component of the gradient to the single-shot acquisition time.
All the steps except for the number of shots are the same as the original iCANS.
In particular, inaccessible quantities $\sigma_i^2$ and $\partial_i f(\bm{\theta}^{(k)})$ are replaced with the bias-corrected exponential moving averages of the previous iterations.
We call this shot allocation strategy Waiting time-Evaluating Coupled Adaptive Number of Shots (weCANS).
To distinguish between iCANS-based version and gCANS-based version (introduced below), we call them weCANS(i) and weCANS(g) respectively.
In this way, we take the explicit value of the latency into account.
Especially, the term proportional to the ratio $R_i$ represents the increase in the number of shots to deal with the latency.
Indeed, as the latency gets large, taking large iteration number becomes more expensive.
Hence, it can be more efficient to take larger number of shots in a single iteration as the latency gets larger compared to $c_1$.
The shot number \eqref{weCANSi_shots} concretely suggests how large number of shots is appropriate based on the estimation of the gain.
For large ratio $R_i$ of the latency, the increase in the number of shots reflecting the latency roughly scales linearly to $\sqrt{R_i L\alpha/(2 - L \alpha)} \sigma_i/\partial_i f(\bm{\theta}^{(k)})$, which implies that the effect of the latency tends to be reflected more for later optimization steps as the ratio of the statistical noise to the signal gets larger in general.

A waiting time-evaluating version of gCANS can also be obtained by maximizing
\begin{align}
 r_{\mathrm{g}}(\bm{s}) = \frac{\mathbb{E}[\mathcal{G}(\bm{s})]}{c_1 \sum_{i=1}^{d} s_i + c_2 \sum_{i=1}^{d} m_i + c_3}.\label{gCANS_max}
\end{align}
This yields the weCANS(g) shot allocation rule
\begin{align}
 s_i = \frac{\sigma_i R}{-\sum_{i=1}^{d}\sigma_i + \sqrt{(\sum_{i=1}^{d}\sigma_i)^2 + R \frac{2 - L\alpha}{L\alpha}\| \bm{\nabla}f(\bm{\theta}^{(k)})\|^2}},\label{weCANSg_shots}
\end{align}
where $R := (c_2 \sum_{i=1}^{d} m_i + c_3) / c_1 $ is the ratio of the overhead per iteration to the single-shot acquisition time.
For weCANS(g), the increase in the number of shots reflecting the latency scales similarly to weCANS(i) as $\sim \sqrt{R L\alpha/(2 - L \alpha)} \sigma_i/\|\nabla f(\bm{\theta}^{(k)})\|$.
Obviously, the leading-order of the gain in the single-shot acquisition time $c_1 r_{\mathrm{g}}(\bm{s})$ is at most $O(1/R)$ for large $R$.
Without latency-aware shot allocation rule in gCANS, the leading-order term of the gain in the single-shot acquisition time is $(\alpha-L\alpha^2/2)\|\bm{\nabla}f(\bm{\theta}^{(k)})\|^2 / (2R)$, whereas using the number of shots of weCANS yields its leading-order term $(\alpha-L\alpha^2/2)\|\bm{\nabla}f(\bm{\theta}^{(k)})\|^2 /R$, which is optimal.
A similar estimation also holds for iCANS.
Hence, roughly speaking, twice acceleration is expected in large $R$ limit.
We remark that subleading-order terms are also maximized in weCANS, although the same leading-order scaling can be obtained by any $s_i \propto R^{a}$ $(0<a<1)$.
Nevertheless, it turns out that such an improvement by weCANS(i) and weCANS(g) was not observed in our numerical simulations as shown in Sec.~\ref{sec_numerics}.
That might be due to the limitation of the estimation of the gain using the expectation value of its lower bound even without considering the statistical error.
On the other hand, incorporating latency-aware adaptive shot strategy into Adam introduced in the next subsection actually works well as shown in Sec.~\ref{sec_numerics}.

\subsection{Waiting time-evaluating CANS for a wider class of SGD including Adam}\label{sec_AdamCANS}

In this section, we propose how to incorporate weCANS adaptive shot number allocation strategy into more general stochastic gradient descent optimizers including Adam.
We consider the following class of the optimizers such that the parameter is updated according to
\begin{align}
 \bm{\theta}^{(k+1)} = \bm{\theta}^{(k)} - \alpha_k \bm{X}_k(\bm{g}(\bm{\theta}^{(k)})),\label{genAdam-eq1}
\end{align}
where $\bm{X}_k(\bm{g}(\bm{\theta}^{(k)}))$ is some analytic function of the gradient estimator.
In Adam, this function is given as
\begin{align}
 \bm{X}_k(\bm{g}(\bm{\theta}^{(k)})) = \bm{m}_{k}(\bm{g}(\bm{\theta}^{(k)})) / \sqrt{\bm{v}_{k}(\bm{g}(\bm{\theta}^{(k)}))},
\end{align}
where $\bm{m}_k(\bm{g}(\bm{\theta}^{(k)}))$ and $\bm{v}_k(\bm{g}(\bm{\theta}^{(k)}))$ are given by Eq.~(\ref{adam_m}) and (\ref{adam_v}) respectively, and we neglect $\epsilon$ in Eq.~(\ref{adam_update}).
Here, we only explicitly consider the dependency of $\bm{X}_k$ on the gradient estimator of the current step.
In particular, $\bm{X}_k$ can depend on the quantities calculated in the previous iterations up to $k - 1$ as in Adam.
The dependency of $\bm{X}_k$ except for $\bm{g}(\bm{\theta}^{(k)})$ is included as the change of the form of the function depending on $k$.
Under this parameter update rule, we have the following lower bound of the absolute value of the gain
\begin{align}
 &|f(\bm{\theta}^{(k)}) - f(\bm{\theta}^{(k+1)})| \nonumber\\
\geq& |\alpha_k \bm{\nabla} f(\bm{\theta}^{(k)})^{\mathrm{T}} \bm{X}_k(\bm{g}(\bm{\theta}^{(k)})[\bm{s}])| - \frac{L\alpha_k^2}{2}\|\bm{X}_k(\bm{g}(\bm{\theta}^{(k)}))[\bm{s}]\|^2 \nonumber\\
=:& \varphi(\bm{g}(\bm{\theta}^{(k)})[\bm{s}])\label{abs_gain_X}
\end{align}
in a similar manner to CANS, where, for clarity, we explicitly denote the dependency on the numbers of shots $\bm{s}$ of the estimator of the gradient
\begin{align}
 {g}_i(\bm{\theta}^{(k)})[\bm{s}] = \frac{1}{s_i} \sum_{k=1}^{s_i} \gamma_{i,k}
\end{align}
with single-shot estimations $\gamma_{i,k}$.
Hence, if we have an estimation of the lower bound $\varphi(\bm{g}(\bm{\theta}^{(k)})[\bm{s}])$, we can determine an appropriate number of shots to maximize the it per unit time.
Here, we consider the absolute value of the gain because the expected update direction given by $\bm{X}_k(\bm{g}(\bm{\theta}^{(k)}))$ can be opposite to the direction of the gradient $\bm{\nabla}f(\bm{\theta}^{(k)})$.
We assume that as large as possible update is beneficial for a given update rule also in such a case.

We remark that the expectation is taken with respect to each single-shot realization $\gamma_{i,k}$ as a random variable.
Hence, we can not obtain an analytic expression for the expectation value of this lower bound in general.
Then, we approximate it via the Taylor expansion of $\varphi$ around the expectation value $\mathbb{E}[\bm{g}(\bm{\theta}^{(k)})[\bm{s}]] = \bm{\nabla}f(\bm{\theta}^{(k)})$ as
\begin{align}
 &\mathbb{E}[\varphi(\bm{g}(\bm{\theta}^{(k)})[\bm{s}])]\nonumber\\
\approx& \varphi(\bm{\nabla}f(\bm{\theta}^{(k)})) + \sum_{i=1}^d \frac{\sigma_i^2}{2s_i}\frac{\partial^2 \varphi}{\partial g_i^2}(\bm{\nabla}f(\bm{\theta}^{(k)}))\nonumber\\
=& A - \sum_{i=1}^{d} \frac{B_i}{s_i},\label{Taylor_adam}
\end{align}
where
\begin{align}
 A :=& \varphi(\bm{\nabla}f(\bm{\theta}^{(k)}))\\
B_i :=& - \frac{\sigma_i^2}{2}\frac{\partial^2 \varphi}{\partial g_i^2}(\bm{\nabla}f(\bm{\theta}^{(k)})).
\end{align}
Eq.~(\ref{Taylor_adam}) has the same form as the objective function for gCANS, except that $B_i$ can be negative.
If $B_i$ is negative, as small as possible value of $s_i$ gives the largest value of our objective function
\begin{align}
 r_{\mathrm{X}}(\bm{s}) = \frac{\mathbb{E}[\varphi(\bm{g}(\bm{\theta}^{(k)})[\bm{s}])]}{c_1 \sum_{i=1}^{d} s_i + c_2 \sum_{i=1}^{d} m_i + c_3}.\label{AdamCANS_max}
\end{align}
to be maximized based on the expression Eq.~(\ref{Taylor_adam}).
However, choosing $s_i = 1$ is not appropriate in this case because Eq.~(\ref{AdamCANS_max}) is based on the Taylor expansion, which is not accurate for small value of $s_i$.
Hence, $B_i \leq 0$ implies that the optimal $s_i$ lies in a region where the Taylor expansion is not accurate.
We circumvent this problem by setting a floor value $s_{\min}$, and we substitute $s_{\min}$ for $s_i$ with $B_i \leq 0$.
In other words, we use $\bm{s}$ given by
\begin{align}
 \bm{s} = \arg \max \{r_{\mathrm{X}}(\bm{s})| s_i \geq s_{\min}\},
\end{align}
which yields the shot allocation rule
\begin{align}
 s_i = 
\begin{cases}
    \frac{(R + |J_{-}| s_{\min})\sqrt{B_i}}{\sqrt{b_{+}^2 + (R + |J_{-}| s_{\min}) (A - \sum_{j\in J_{-}} \frac{B_j}{s_{\min}})} - b_{+}} & \text{if $B_i > 0$,} \\
    s_{\min}       & \text{if $B_i \leq 0$,}
 \end{cases}\label{shots_weAdam}
\end{align}
where $R = (c_2 \sum_{i=1}^{d} m_i + c_3) / c_1 $ is the ratio of the overhead per iteration to the single-shot acquisition time and
\begin{align}
 b_{+} := \sum_{j\in J_{+}} \sqrt{B_j}
\end{align}
with $J_{+}:=\{j| B_j \geq 0\}$ and $J_{-}:=\{j| B_j < 0\}$.
All the inaccessible quantities are also replaced with the exponential moving average calculated from the previous iterations in the same way as iCANS and gCANS.
For Eq.~(\ref{shots_weAdam}) to give valid numbers of shots, $A>0$ is required, which is equivalent to imposing
\begin{align}
 \alpha_k < \frac{2 |\bm{\nabla}f(\bm{\theta}^{(k)})^{\mathrm{T}} \bm{X}(\bm{\nabla}f(\bm{\theta}^{(k)}))|}{L \|\bm{X}(\bm{\nabla}f(\bm{\theta}^{(k)}))\|^2}\label{alpha_X_cond}
\end{align}
in accordance with the condition for the lower bound (\ref{abs_gain_X}) to be positive.
We can forcibly impose Eq.~(\ref{alpha_X_cond}) by clipping $\alpha_k$ by $r$ times this upper limit with a fixed constant $0 < r <1$.
However, in practice, clipping $\alpha_k$ may delay the optimization.
Instead, we can use a heuristic strategy to use the original $\alpha_k$ for the parameter update while the clipped value is used to calculate the number of shots.
We take this strategy in the numerical simulations in Sec.~\ref{sec_numerics}.
\begin{figure}[!t]
\begin{algorithm}[H]
    \caption{Adam with weCANS adaptive shot strategy (we-AdamCANS). The function $iEvaluate(\bm{\theta}, \bm{s})$ evaluates the gradient at the parameter $\bm{\theta}$ using $s_i$ shots to estimate $i$-th partial derivative via parameter-shift rule (\ref{PSR_2}) or (\ref{GPSR}). This function returns the vectors $\bm{g}$ and $\bm{S}$ whose components are the estimated partial derivatives and its variance respectively. The function $ShotsEvaluate(A, \bm{B}, R, s_{\min})$ evaluates the number of shots via Eq.~(\ref{shots_weAdam}) using given $A$, $\bm{B}$, $R$, and $s_{\min}$, and returns the vector $\bm{s}$ of the integers.}
    \label{alg1}
    \begin{algorithmic}[1]
    \REQUIRE Step size $\alpha$, initial parameter $\bm{\theta}_0$, minimum number of shots $s_{\min}$, total wall-clock time $T$ available for the optimization, hyperparameters $\beta_1$, $\beta_2$, and $\epsilon$, Lipschitz constant $L$, running average constant $\mu$, ratio $R$ of the overhead to the single-shot acquisition time, step-size clipping rate $r$, Boolean flag $Clipping$
    \STATE Initialize: $\bm{\theta} \leftarrow \bm{\theta}_0$, $t \leftarrow 0$, $t_{\mathrm{start}} \leftarrow$ current wall-clock time, $\bm{s} \leftarrow (s_{\min},\cdots,s_{\min})^{\mathrm{T}}$, $\bm{\chi}' \leftarrow (0,\cdots,0)^{\mathrm{T}}$, $\bm{\xi}' \leftarrow (0,\cdots,0)^{\mathrm{T}}$, $\bm{m}' \leftarrow (0,\cdots,0)^{\mathrm{T}}$, $\bm{v}' \leftarrow (0,\cdots,0)^{\mathrm{T}}$, $k \leftarrow 0$, $\alpha' \leftarrow \alpha$
    \WHILE{$t \leq T$}
    \STATE Evaluate the current wall-clock time $t_{\mathrm{start}}$
    \STATE $\bm{g}, \bm{S} \leftarrow iEvaluate(\bm{\theta}, \bm{s})$
    \STATE $\bm{m}' \leftarrow \beta_1 \bm{m}' + (1 - \beta_1) \bm{g}$
    \STATE $\bm{v}' \leftarrow \beta_2 \bm{v}' + (1 - \beta_2) \bm{g}^2$
    \STATE $\bm{m} \leftarrow \bm{m}'/(1 - \beta_1^k)$
    \STATE $\bm{v} \leftarrow \bm{v}'/(1 - \beta_2^k)$
    \STATE $\bm{\theta} \leftarrow \bm{\theta} - \alpha \bm{m} / (\sqrt{\bm{v}}+\epsilon)$
    \STATE $\bm{\chi}' \leftarrow \mu \bm{\chi}' + (1 - \mu)\bm{g}$
    \STATE $\bm{\xi}' \leftarrow \mu \bm{\xi}' + (1 - \mu)\bm{S}$
    \STATE $\bm{\xi}' \leftarrow \bm{\xi}' / (1 - \mu^{k+1})$
    \STATE $\bm{M} \leftarrow \beta_1 \bm{m}' + (1 - \beta_1) \bm{\chi}$
    \STATE $\bm{V} \leftarrow \beta_2 \bm{v}' + (1 - \beta_2) \bm{\chi}^2$
    \STATE $\bm{M} \leftarrow \bm{M}/(1 - \beta_1^k)$
    \STATE $\bm{V} \leftarrow \bm{V}/(1 - \beta_2^k)$
    \STATE $\bm{X} \leftarrow \bm{M} / (\sqrt{\bm{V}}+\epsilon)$
    \STATE $\alpha \leftarrow \min\{\alpha', r\frac{2 |\bm{\chi}^{\mathrm{T}} \bm{X}|}{L \|\bm{X}\|^2}\}$
    \STATE $A \leftarrow \varphi(\bm{\chi})$ using $\alpha$ as the step size in Eq.~(\ref{abs_gain_X})
    \FOR{$i\in [1,\cdots, d]$}
    \STATE $B_i \leftarrow - \frac{\xi_i}{2}\frac{\partial^2 \varphi}{\partial g_i^2}(\bm{\chi})$
    \ENDFOR
    \STATE $\bm{s} \leftarrow ShotsEvaluate(A, \bm{B}, R, s_{\min})$
    \IF{not $Clipping$}
    \STATE $\alpha \leftarrow \alpha'$
    \ENDIF
    \STATE $t_{\mathrm{end}} \leftarrow$ current wall-clock time
    \STATE $t \leftarrow t + t_{\mathrm{end}} - t_{\mathrm{start}}$
    \STATE $t_{\mathrm{start}} \leftarrow t_{\mathrm{end}}$
    \STATE $k \leftarrow k + 1$
    \ENDWHILE
    \end{algorithmic}
\end{algorithm}
\end{figure}
We also call this adaptive shot allocation strategy weCANS.
Especially, we call the Adam optimizer using weCANS strategy we-AdamCANS.
For clarity, we describe the detail of we-AdamCANS in Algorithm \ref{alg1}.

We also obtain the shot allocation rule without considering the latency by taking the limit $R\rightarrow 0$.
If $|J_{-}|=0$, we have
\begin{align}
 s_i = 2\frac{\sqrt{B_i}\sum_{j=1}^d \sqrt{B_j}}{A}.
\end{align}
Especially, we call the Adam optimizer using this adaptive number of shots AdamCANS.

The best expected gain in the single-shot acquisition time of the above generalized weCANS is also twice as large as that of generalized CANS without latency consideration in large $R$ limit, in the same way as the weCANS(g) (weCANS(i)) vs gCANS (iCANS).
As shown in numerical simulations in Sec.~\ref{sec_numerics}, we-AdamCANS actually achieves fast convergence in wall-clock time, although its acceleration is not always twice in comparison with AdamCANS (sometimes less than twice, sometimes even more faster).

\subsection{Incorporating WRS into weCANS}\label{sec_weCANS-WRS}

For $H$ composed of multiple noncommuting observables, we allocate the number of shots for each simultaneously measurable group according to WRS as explained in Sec.~\ref{sec_op_sampling}.
In this case, the number of circuits $m_i$ used to estimate the gradient is also random because the latency only occurs for the measured groups.
Hence, we cannot access $m_i$ before we decide the number of shots.
Then, we take the following heuristic way to obtain an approximated maximization of the gain per unit time $r_{o}$ $(o = \mathrm{i}, \mathrm{g}, \mathrm{X})$ by replacing $m_i$ with an accessible quantity.
For simplicity, we consider the case where the two-point parameter shift rule (\ref{PSR_2}) holds and $s_i/2$ shots are used for the evaluation of the cost function at each shifted parameter.
It is straightforward to generalize the following prescription to more general case with Eq.~(\ref{GPSR}).
We assume that we have obtained a grouping of the Pauli observables in Eq.~(\ref{pauli_decomp}) into $M$ groups of simultaneously measurable observables as
\begin{align}
 H = \sum_{j=1}^{M}\sum_{i\in G_j} c_i P_i.
\end{align}
Then, each group $G_j$ is measured with probability $p_j = \sum_{i\in G_j}|c_i| / \sum_{k=1}^N |c_k|$ in WRS.
We remark that we can use any probability distribution $p_j$ in general.
Hence, group $G_j$ is not measured with probability $(1 - p_j)^{s_i/2}$ in each evaluation of the cost function.
Thus, the expectation value of $m_i (s_i)$ when $s_i$ shots are used is
\begin{align}
 \mathbb{E}[m_i(s_i)] = 2 \left(M - \sum_{j=1}^{M} (1 - p_j)^{s_i/2}\right).\label{m_i_expect}
\end{align}
However, this value still depends on $s_i$.
Although it may be possible to numerically solve the maximization problem of $r_{o}$ with $m_i$ replaced with its expectation value (\ref{m_i_expect}) which we denote by $\tilde{r}_{o}$ from scratch, it might be expensive.
In such a case, we instead use the exponential moving average $\tilde{s}_{i}$ of $s_i$ calculated from the previous iterations in place of $s_i$.
We can further neglect the terms $(1 - p_j)^{\tilde{s}_i/2}$ below some small value since the probability of no occurrence of such a term is too small, and it is expected to be better to count its latency.
Then, we obtain the analytic expressions of $s_i$ for weCANS (\ref{weCANSi_shots}) and (\ref{weCANSg_shots}) in the same way.
Moreover, we can further implement a full maximization of $\tilde{r}_{o}$ by using the obtained analytic expressions as the initial guess.

\section{Numerical simulations}\label{sec_numerics}
In this section, we demonstrate the performance of weCANS optimizers in comparison with other state-of-the-art optimizers through numerical simulations of three kinds of benchmark VQA problems.
We compare weCANS(i), weCANS(g), we-AdamCANS, AdamCANS with the original iCANS, gCANS, Adam with some fixed number of shots, and SHOALS.
Including the weCANS version, $1 / L$ is used as the step size for iCANS and gCANS, the other hyperparameters are the same as the default values of iCANS and gCANS.
$\beta_1=0.9$, $\beta_2 = 0.99$, $\epsilon = 10^{-8}$, and $\alpha = 1/ L$ are used in we-AdamCANS, AdamCANS, Adam.
As for the other specific hyperparameters of we-AdamCANS and AdamCANS, $s_{\min} = 100$, $\mu=0.99$, and $r=0.75$ are used.
We employ $Clipping = \mathrm{False}$ in Algorithm \ref{alg1}.

For the other hyperparameters, the default values of each optimizer given in each proposal paper are used.
SHOALS sometimes gets stuck in too small values of the step size in the line search.
To avoid this, we introduced the minimum of the step size as $10^{-4}$.
We implement all the simulations via Qulacs \cite{Suzuki2021qulacsfast}.
\subsection{Compiling task}
\begin{figure}
\centering
 \includegraphics[clip ,width=3.2in]{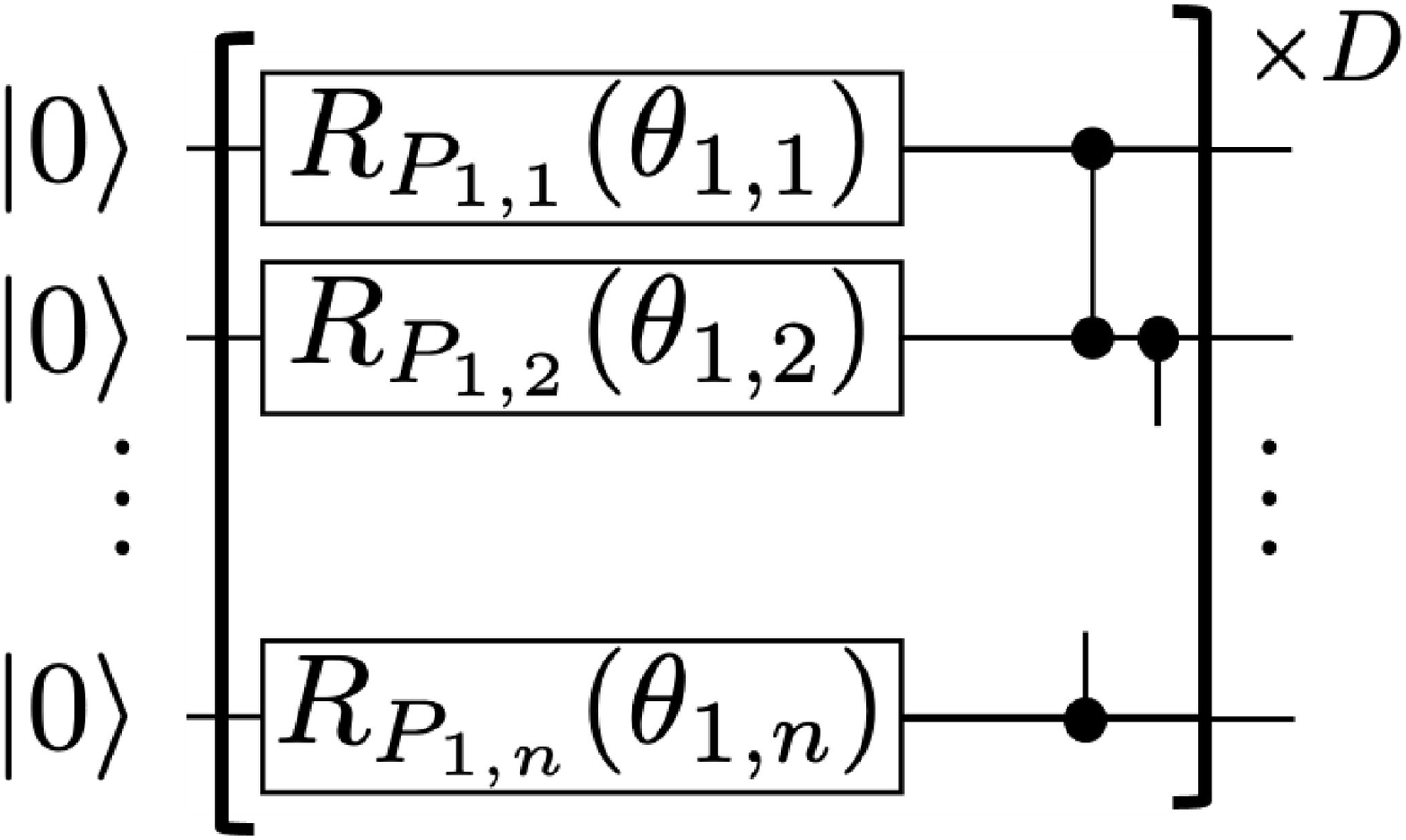}
 \caption{The parameterized quantum circuit used in the compiling task. Each layer of the circuit is composed of single-qubit Pauli rotation with a randomly chosen axis of each qubit followed by the controlled-$Z$ gates on the adjacent qubits.}
\label{fig-circuit}
\end{figure}
\begin{figure}
\centering
 \includegraphics[clip ,width=3.2in]{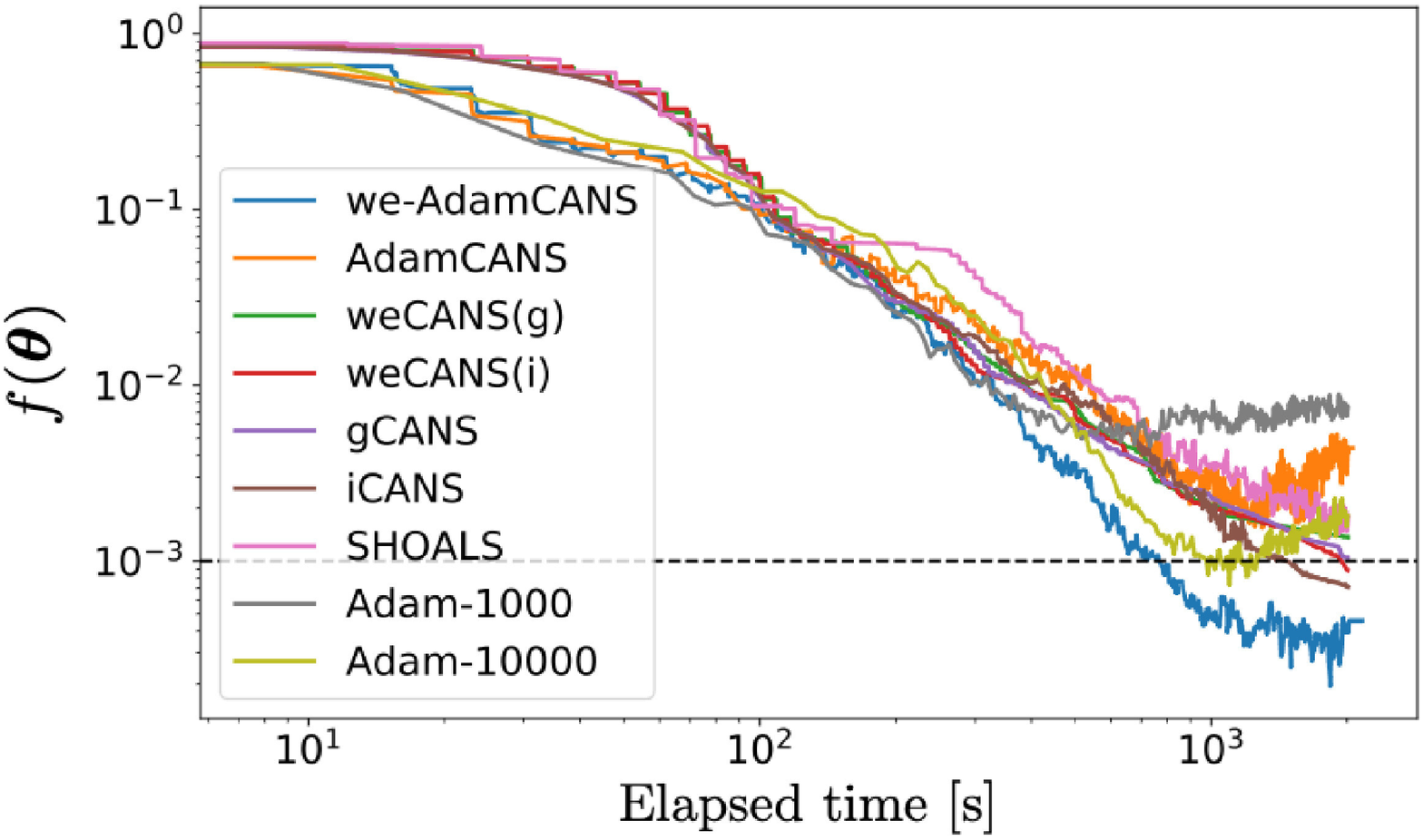}
 \caption{Comparison of performance of the optimizers for the compilation benchmark. The dashed line indicates the precision $10^{-3}$. Adam-$s$ means Adam using $s$ shots for every evaluation of the cost function. The median of the exact values of the cost function at the parameters obtained in 30 trials of the optimizations is shown vs the elapsed time with the single-shot time $c_1 = 1.0 \times 10^{-5}$ s, circuit switching latency $c_2 = 0.1$ s, and the communication latency $c_3 = 4.0$ s.}
\label{fig-opt}
\end{figure}
As a first benchmark problem, we consider a variational compilation task \cite{nakanishi_sequential_2020,Kubler2020adaptiveoptimizer}.
Our task is to minimize the infidelity of the state $U(\bm{\theta})\ket{\bm{0}}$
\begin{align}
 f(\bm{\theta}) = 1 - |\bra{\bm{0}} U(\bm{\theta}^{*})^{\dagger}U(\bm{\theta})\ket{\bm{0}}|^2
\end{align}
with the target state $U(\bm{\theta}^{*})^{\dagger} \ket{\bm{0}}$ given by a randomly selected parameter vector $\bm{\theta}^{*}$ as the cost function.
In this way, the ansatz always has the exact solution at the target parameter.

The parameterized quantum circuit $U(\vec{\theta})$ used in the numerical simulation is shown in Fig.~\ref{fig-circuit}, where $R_{P_{i,j}}(\theta_{i,j}) = \exp[- i \theta_{i,j} P_{i,j}/2]$ with $P_{i,j}$ uniformly and randomly selected from the Pauli operators $X, Y, Z$.
Especially, we simulate the optimization with $n=3$ qubits and $D=3$ depth.

The result of our simulation is shown in Fig.~\ref{fig-opt}.
We use the latency values $c_1 = 1.0 \times 10^{-5}$ s, $c_2 = 0.1$ s, and $c_3 = 4.0$ s following \cite{Sung_2020,2201.13438}.
Adam-$s$ means Adam using $s$ shots for every evaluation of the cost function.
The median of the exact values of the cost function $f(\bm{\theta})$ at the optimized parameters of 30 runs is shown as the metric of the performance.
Actually, we-AdamCANS outperforms the other optimizers.
Especially, we-AdamCANS attains the precision $1.0\times 10^{-3}$ in $773$ s whereas Adam-10000 attains the same precision at 985 s but the error increases again due to the statistical error.
The next faster one is iCANS which attains the same precision in 1463 s.
Hence, we-AdamCANS is over about 2 times faster than the other optimizers in this problem.
However, no improvement can be seen in the median performance by the simple strategies weCANS(i) and weCANS(g).
That might be because our estimations are based on the maximization of the expectation value of the lower bound of the gain but not the gain itself.
Hence the performance may be improved by using a tighter estimation of the Lipschitz constant $L$ which yields tighter estimation of the lower bound of the gain.


\subsection{VQE tasks of quantum chemistry}
\begin{figure}
\centering
 \includegraphics[clip ,width=3.2in]{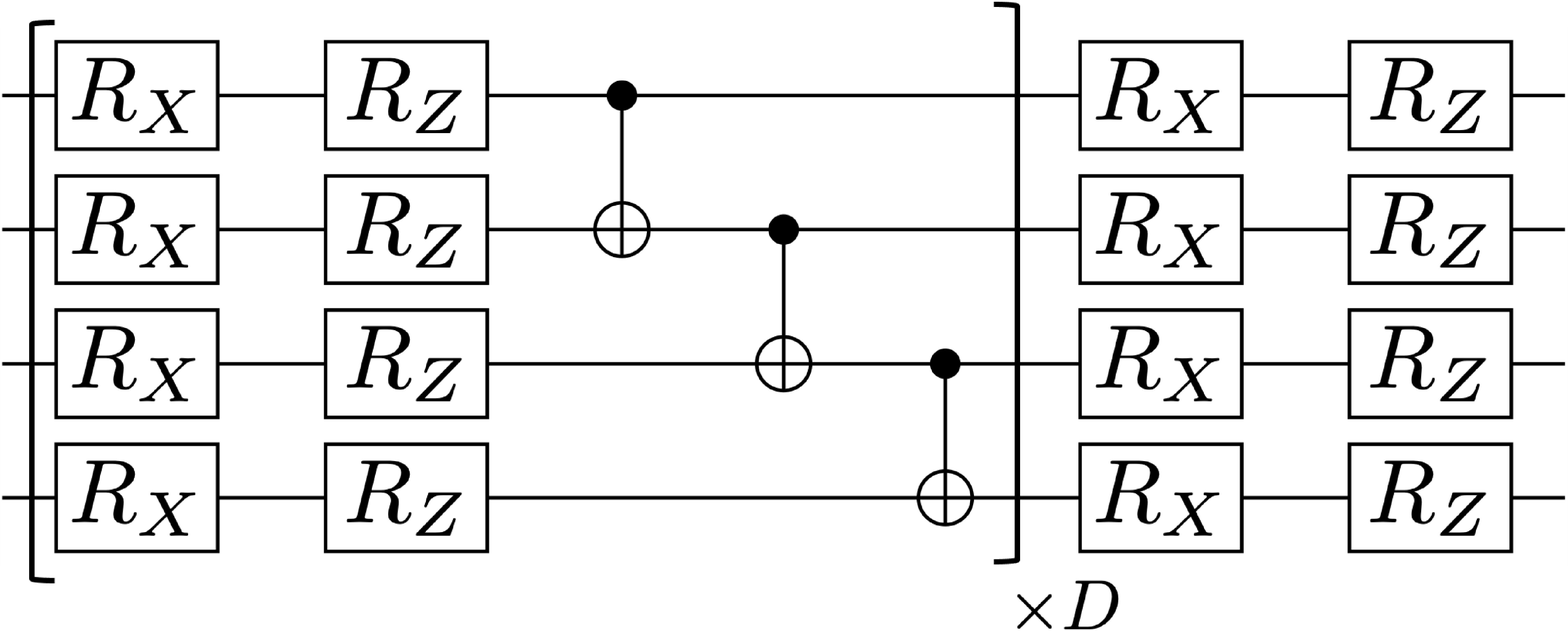}
 \caption{The hardware efficient ansatz used in the numerical simulation of VQE tasks of quantum chemistry. Each layer of the circuit is composed of single-qubit $X$ rotation and $Z$ rotation of each qubit followed by the controlled-not gates on the adjacent qubits.}
\label{fig-circuit_HEA_VQE}
\end{figure}
\begin{figure*}
\centering
 \includegraphics[clip ,width=7.2in]{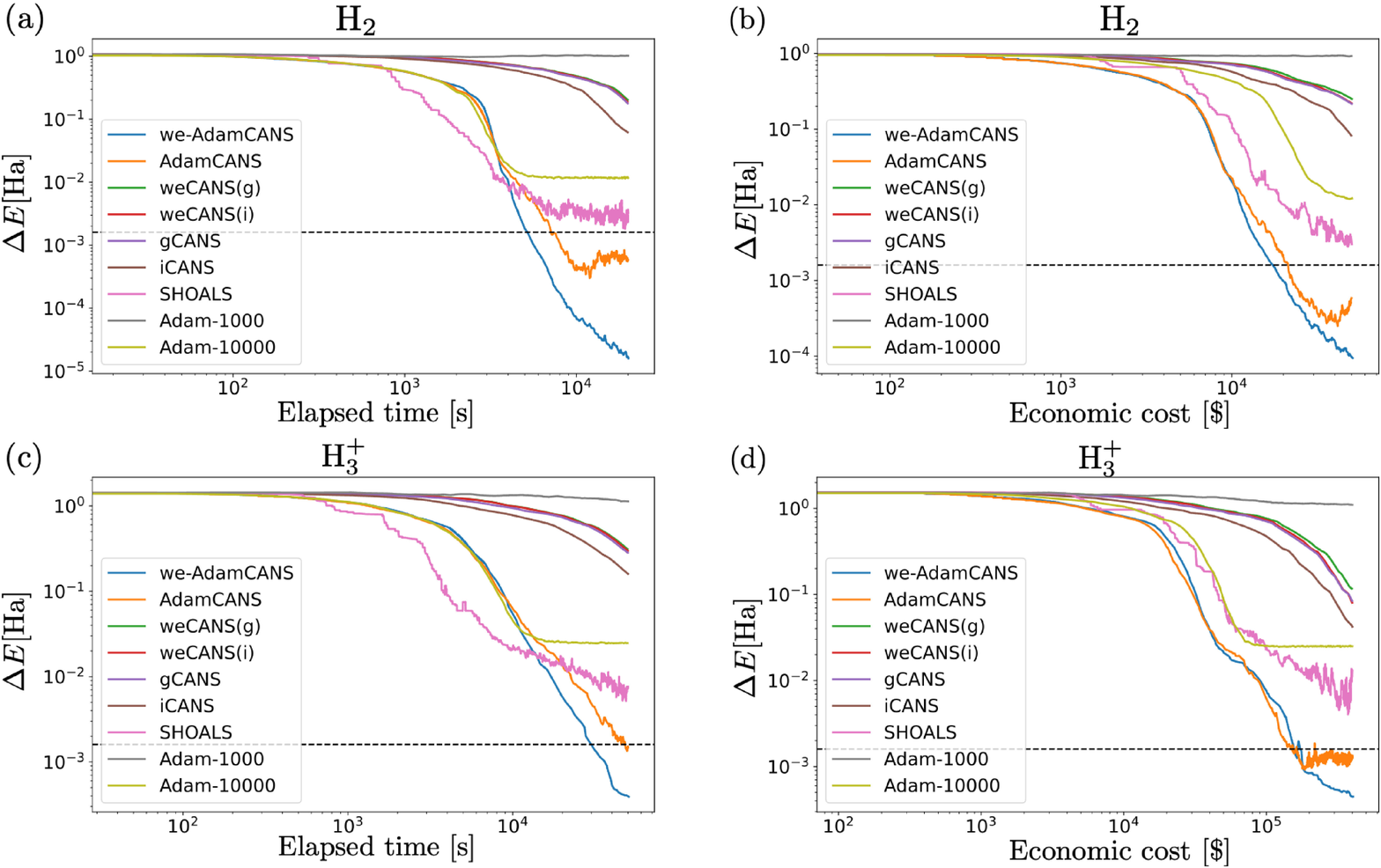}
 \caption{Comparison of performance of the optimizers for the VQE benchmark. Every graph shows the median of 30 trials of the optimization. The dashed line indicates the chemical accuracy 1.6$\times 10^{-3}$ Ha. (a) $\Delta E$ vs the elapsed time for $\mathrm{H}_2$ with $c_1=10^{-5}$ s, $c_2=0.1$ s and $c_3=4.0$ s. (b) $\Delta E$ vs the economic cost for $\mathrm{H}_2$ with the pricing of the Rigetti's devices in Amazon Braket where $c_1=\text{\$}3.5\times 10^{-4}$, $c_2=\text{\$}0.3$. (c) $\Delta E$ vs the elapsed time for $\mathrm{H}_3^{+}$ with $c_1=10^{-5}$ s, $c_2=0.1$ s and $c_3=4.0$ s (d) $\Delta E$ vs the economic cost for $\mathrm{H}_3^{+}$ with the pricing of the Rigetti's devices in Amazon Braket where $c_1=\text{\$}3.5\times 10^{-4}$, $c_2=\text{\$}0.3$.}
\label{fig-VQE_chem}
\end{figure*}
\begin{figure*}
\centering
 \includegraphics[clip ,width=7.2in]{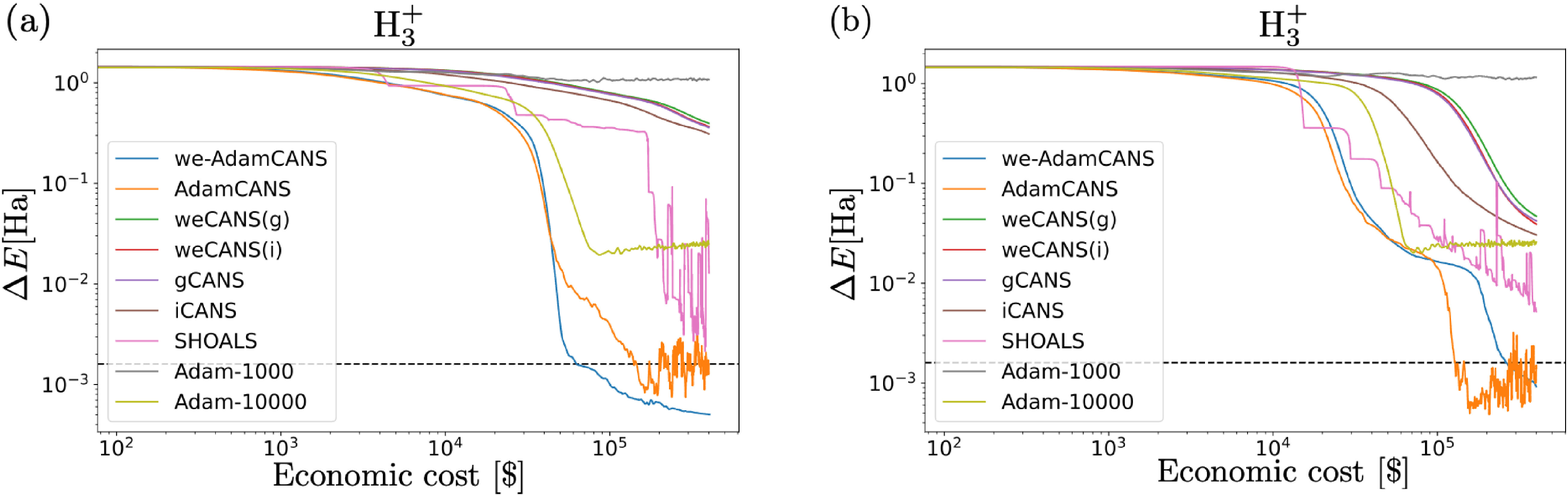}
 \caption{Comparison of performance in a single run of the optimizers for the VQE benchmark for $\mathrm{H}_3^{+}$ in terms of the economic cost with the pricing of the Rigetti's devices in Amazon Braket. (a) and (b) show two examples of a single run of the optimization out of 30 trials.}
\label{fig-single_run_VQE_chem}
\end{figure*}
\begin{figure*}
\centering
 \includegraphics[clip ,width=7.2in]{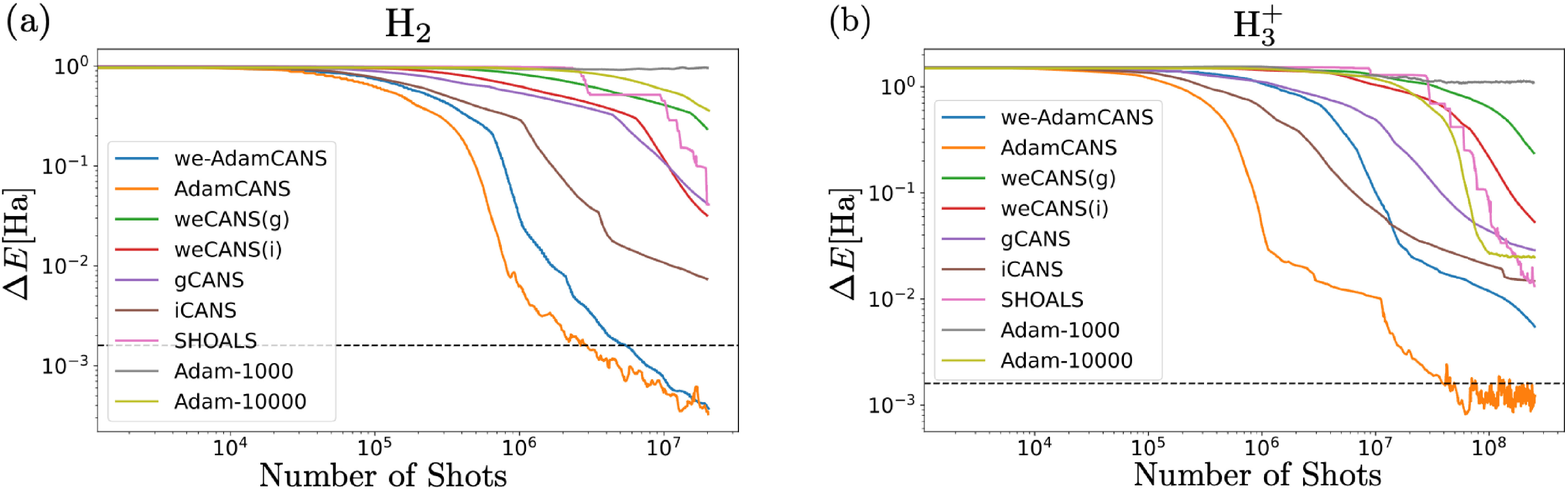}
 \caption{Comparison of performance of the optimizers for the VQE benchmark in terms of the number of shots. Both graphs show the median of 10 trials of the optimization. The dashed line indicates the chemical accuracy 1.6$\times 10^{-3}$ Ha. (a) $\Delta E$ vs the number of shots used for $\mathrm{H}_2$ (b) $\Delta E$ vs the number of shots used for $\mathrm{H}_3^{+}$. For weCANS optimizers, we input the latency values $c_1=3.0\times 10^{-4}$, $c_2=0.3$ and $c_3=0.0$ for (a), and $c_1=10^{-5}$, $c_2=0.1$ and $c_3=4.0$ for (b) to calculate the number of shots.}
\label{fig-shots_VQE_chem}
\end{figure*}

As the next benchmark problem, we consider VQE problems of quantum chemistry.
We consider $\mathrm{H}_2$ molecule and $\mathrm{H}_3^{+}$ chain as examples.
The distance between H atoms is set to 0.74\AA\: for both molecules.
For $\mathrm{H}_2$ molecule, we use the STO-3G basis to encode the molecular Hamiltonian into the qubit systems via Jordan-Wigner mapping.
For a benchmark purpose, we consider the VQE task of $\mathrm{H}_2$ encoded in $4$ qubits in this way without any qubit reduction.
For $\mathrm{H}_3^{+}$ molecule, we also use the STO-3G basis, but apply the parity transform \cite{1208.5986} as the fermion-to-qubit mapping.
We further reduce two qubits via $\mathbb{Z}_2$ symmetry.
Then, we obtain a 4-qubit Hamiltonian for $\mathrm{H}_3^{+}$ molecule.
We just use a greedy algorithm to group the Pauli terms into qubit-wise commutable terms, where we use the coefficient of each term as the weight.
We use a hardware efficient ansatz shown in Fig.~\ref{fig-circuit_HEA_VQE} as the ansatz with $D=3$.

Fig.~\ref{fig-VQE_chem} shows the numerical results for the simulation of VQE.
Fig.~\ref{fig-VQE_chem}(a) and (c) respectively shows the performance of the optimizers for VQE tasks of $\mathrm{H}_2$ and $\mathrm{H}_3^{+}$ in terms of the elapsed time under the single-shot time $c_1=10^{-5}$ s, circuit-switching latency $c_2=0.1$ s, and the communication latency $c_3=4.0$ s, where $\Delta E$ denotes the difference of the exact energy expectation value evaluated at the parameters obtained by the optimization from the ground-state energy.
On the other hand, Fig.~\ref{fig-VQE_chem}(b) and (d) respectively shows the performance of the optimizers for VQE tasks of $\mathrm{H}_2$ and $\mathrm{H}_3^{+}$ in terms of the economic cost for using the Rigetti's devices in Amazon Braket where $c_1=\text{\$}3.5\times 10^{-4}$, $c_2=\text{\$}0.3$ and $c_3 = \text{\$}0.0$.
We remark that the economic cost can be relabeled as the elapsed time, which shows the performance without the communication latency.
In all the cases, we-AdamCANS considerably outperforms the other optimizers except for AdamCANS.
In particular, only we-AdamCANS and AdamCANS attain the chemical accuracy 1.6$\times 10^{-3}$ Ha within the range of the simulation ($2\times 10^4$ s and $\text{\$}5\times 10^4$ for $H_2$, $5\times 10^4$ s and $\text{\$}4\times 10^5$ for $H_3^{+}$).
On the other hand, the simple strategies weCANS(i) and weCANS(g) do not work well again in this case.

Except for the VQE performance for $\mathrm{H}_3^{+}$ in terms of the economic cost, latency-aware shot allocation strategy of we-AdamCANS actually yields faster convergence than AdamCANS without latency consideration.
However, in that exceptional case of $\mathrm{H}_3^{+}$ VQE vs the economic cost, we-AdamCANS and AdamCANS attains the chemical accuracy almost with the same cost, though the convergence of we-AdamCANS is more clear and further precision is achieved.
That seems because we-AdamCANS sometimes happens to delay by being trapped on a plateau in this problem.
In fact, we can verify this observation by looking at some examples of a single run of the optimization shown in Fig.~\ref{fig-single_run_VQE_chem}.
In the case of Fig.~\ref{fig-single_run_VQE_chem} (a), we-AdamCANS actually achieves considerably faster convergence than AdamCANS.
On the other hand, in the case of Fig.~\ref{fig-single_run_VQE_chem} (b), both we-AdamCANS and AdamCANS seems trapped on a plateau.
In this case, AdamCANS escapes from the plateau faster than we-AdamCANS.
This phenomenon can be understood from the fact that fluctuations can assist to escape from a plateau.
Because we-AdamCANS uses larger number of shots to taking into account the latency, AdamCANS can have more chance to escape from a plateau thanks to its larger fluctuations.
In the case of the larger ratio of the overhead in Fig.~\ref{fig-VQE_chem} (c),
the delay of we-AdamCANS due to the plateau might be compensated by the delay of AdamCANS due to the large latency.
It is left for future works to further improve we-AdamCANS shot allocation strategy in such a way that pros and cons of fluctuations are appropriately balanced
to avoid such a plateau problem.

Next, let us also see the performance in terms of the number of shots.
Fig.~\ref{fig-shots_VQE_chem} shows the performance of the optimizers in terms of the number of shots used.
Fig.~\ref{fig-shots_VQE_chem} (a) and (b) are the results for $\mathrm{H}_2$ and $\mathrm{H}_3^{+}$ respectively.
For weCANS optimizers, we used the latency values $c_1=3.0\times 10^{-4}$, $c_2=0.3$ and $c_3=0.0$ for (a), and $c_1=10^{-5}$, $c_2=0.1$ and $c_3=4.0$ for (b) to calculate the number of shots.
If we measure the performance in terms of the total expended shots, AdamCANS indeed outperforms the other optimizers in both problems.

\subsection{VQE of 1-dimensional Transverse field Ising model}\label{sec_TFIM}
\begin{figure}
\centering
 \includegraphics[clip ,width=3.2in]{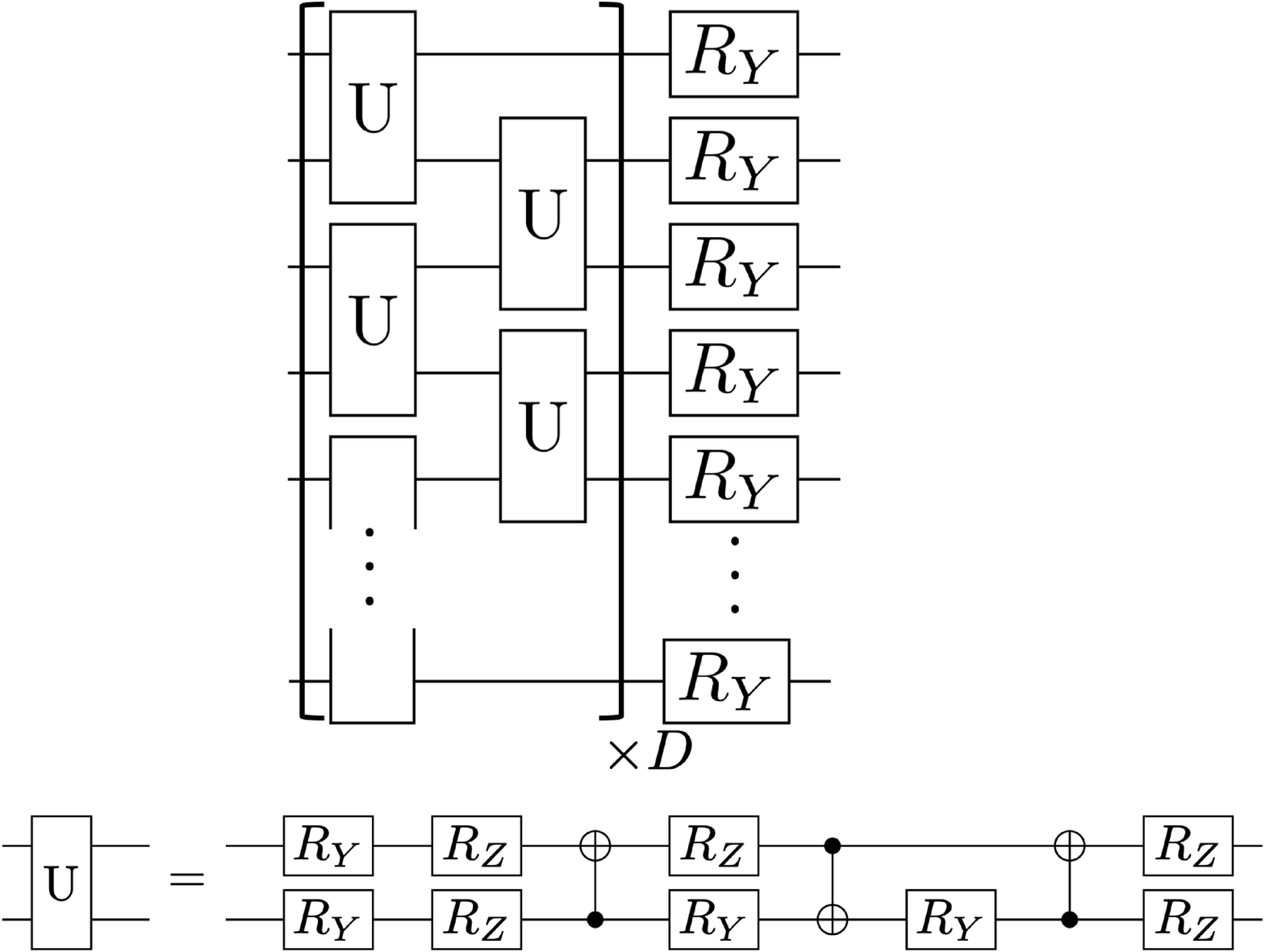}
 \caption{The parameterized quantum circuit used in the numerical simulation of the VQE task of a 1-dimensional transverse Ising spin chain.}
\label{fig-circuit-TFIM}
\end{figure}
\begin{figure*}
\centering
 \includegraphics[clip ,width=7.2in]{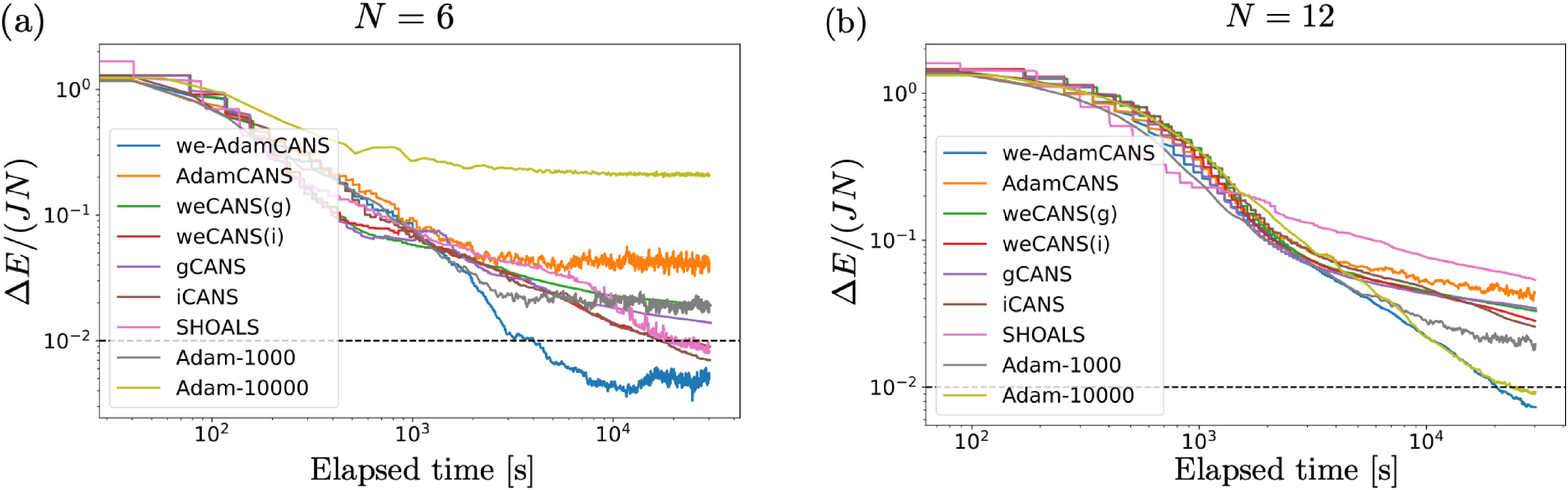}
 \caption{Comparison of performance of the optimizers for the VQE tasks of 1d transverse Ising spin chain with open boundary conditions. The median of the exact value of the energy difference per site $\Delta E / (J N)$ at the parameters obtained by each optimizer in 10 runs of the optimizations is shown vs the elapsed time with the single-shot time $c_1 = 1.0 \times 10^{-5}$ s, circuit switching latency $c_2 = 0.1$ s, and the communication latency $c_3 = 4.0$ s for (a) $N=6$ (b) $N=12$. The dashed line indicates the precision $10^{-2}$.}
\label{fig-opt-TFIM}
\end{figure*}

Finally, in order to demonstrate the performance of our optimizers in larger system sizes, we consider the VQE task of the 1-dimensional transverse field Ising spin chain with open boundary conditions for the system size $N=6$ and 12, where the Hamiltonian is
\begin{align}
 H = -J\sum_{i=1}^{N-1}Z_i Z_{i+1} - g\sum_{i=1}^{N} X_i.
\end{align}
We consider the case with $g/J = 1.5$.
For this problem, we used the ansatz shown in Fig.~\ref{fig-circuit-TFIM} with $D=3$ following \cite{Kubler2020adaptiveoptimizer}.
The ratio of the number of shots for the measurements of the interaction term to that of the transverse field term is deterministically fixed as $J$ to $g$.

Fig.~\ref{fig-VQE_chem}(a) and (b) respectively shows the performance of the optimizers for VQE tasks for $N=6$ and $N=12$ in terms of the elapsed time under the single-shot time $c_1=10^{-5}$ s, circuit-switching latency $c_2=0.1$ s, and the communication latency $c_3=4.0$ s.
The results are shown in terms of the precision of the per-site energy $\Delta E / (J N)$, where $\Delta E$ denotes the difference of the exact energy expectation value evaluated at the parameters obtained by the optimization from the ground-state energy.
Actually, we-AdamCANS outperforms the other implemented optimizers for both $N=6$ and $N=12$.
In particular, we-AdamCANS attains the precision $10^{-2}$ considerably faster than the other optimizers except for Adam-10000 for $N=12$.
However, the performance of Adam with a fixed number of shots is sensitive to the choice of the number of shots.
In fact, Adam-10000 is too slow for $N=6$.
On the other hand, we emphasize that there is no need to tune the number of shots to use in we-AdamCANS, which indeed robustly achieves fast convergence for both systems sizes.
Especially, for $N=6$, we-AdamCANS attains $10^{-2}$ in 4044 s whereas the next fastest iCANS attains the same precision in 17455 s.
Hence, we-AdamCANS is more than about 4-times faster than the others for the convergence up to this precision in this case.

\section{Conclusion}
\label{jzbmlzhd}
In this study, we proposed waiting-time evaluating strategy, weCANS, for determining the number of shots for SGD optimizers in variational quantum algorithms to enhance convergence in terms of wall-clock time or cost for using a cloud service.
Our weCANS strategy is applicable to a wide range of variations of SGD, including Adam.
As shown in numerical simulations, Adam with weCANS (we-AdamCANS) actually outperforms existing shot-adaptive optimizers in terms of convergence speed and cost, as well as shot count in the absence of latency.
On the other hand, we have also shown that just applying weCANS to the simple SGD (weCANS(i) and weCANS(g)) is less effective, possibly because the gain estimation is too rough for them.

While our results demonstrate the effectiveness of we-AdamCANS, there is still room for improvement. For example, considering the positive impact of statistical fluctuations on escaping plateaus may overcome a potential drawback of weCANS that the risk of being trapped in a plateau is increased due to its increased number of shots.
In addition, optimizing shot allocation for each term in the Hamiltonian may lead to further performance enhancements. Additionally, generalizing the adaptive shot strategy to a wider range of optimizers and exploring its practical application in VQAs are open avenues for future research.
We also remark that reducing the latency time is crucial in improving the overall performance of variational quantum algorithms since the maximum expected gain in single-shot acquisition time inevitably scales as $O(1/R)$ with the overhead ratio $R$ by using any shot allocation strategies.
\begin{acknowledgments}
The author would like to thank Keisuke Fujii and Wataru Mizukami for very helpful comments, and thank Kosuke Mitarai for his useful python modules.
This work is supported by MEXT Quantum Leap Flagship Program (MEXT QLEAP) Grant Number JPMXS0118067394 and JPMXS0120319794.
\end{acknowledgments}


%

\end{document}